\documentclass[amsmath,amssymb,aps,prd,11pt,tightenlines,superscriptaddress,nofootinbib,preprintnumbers,notitlepage]{revtex4-1}

\newcommand{\myaffiliation}[1]{\affiliation{#1}}

\input{header}

\begin{document}


\pagestyle{empty}

\vspace*{-0.1in}
\title{{\large THE FORWARD PHYSICS FACILITY AT THE \\LARGE HADRON COLLIDER} \\
 \vspace*{1.88in} 
{\mbox{\normalsize \normalfont on behalf of the FPF Working Groups} \\
\vspace*{0.15in}
{\small \normalfont
Contact Information: alan.barr@physics.ox.ac.uk, jamie.boyd@cern.ch, 
Albert.de.Roeck@cern.ch, } \\
{\small \normalfont
felix.kling@desy.de, joshua.angus.mcfayden@cern.ch, j.rojo@vu.nl} }
\vspace*{-2.0in} 
}


\author{Luis~A.~Anchordoqui}
\myaffiliation{Department of Physics and Astronomy, Lehman College, City University of New York, Bronx, NY 10468, USA}

\author{Akitaka Ariga}
\myaffiliation{Albert Einstein Center for Fundamental Physics, Laboratory for High Energy Physics, \\ University of Bern, Sidlerstrasse 5, CH-3012 Bern, Switzerland}
\myaffiliation{\mbox{Department of Physics, Chiba University, 1-33 Yayoi-cho Inage-ku, Chiba, 263-8522, Japan}}

\author{Tomoko Ariga}
\myaffiliation{Kyushu University, Nishi-ku, 819-0395 Fukuoka, Japan}

\author{Alan~J.~Barr}
\myaffiliation{Department of Physics, University of Oxford, OX1 3RH, United Kingdom}

\author{Brian Batell}
\myaffiliation{Department of Physics and Astronomy, University of Pittsburgh, Pittsburgh, PA 15217, USA}

\author{Jianming Bian}
\myaffiliation{Department of Physics and Astronomy, University of California, Irvine, CA 92697-4575, USA}

\author{Jamie Boyd}
\myaffiliation{CERN, CH-1211 Geneva 23, Switzerland}

\author{Matthew Citron}
\myaffiliation{\mbox{Department of Physics and Astronomy, University of California, Davis, CA 95616, USA}}

\author{Albert De Roeck}
\myaffiliation{CERN, CH-1211 Geneva 23, Switzerland}

\author{Milind~V.~Diwan}
\myaffiliation{Brookhaven National Laboratory, Upton, NY 11973, USA}

\author{Jonathan~L.~Feng}
\myaffiliation{Department of Physics and Astronomy, University of California, Irvine, CA 92697-4575, USA}

\author{Christopher~S.~Hill}
\myaffiliation{Department of Physics, The Ohio State University, Columbus, OH 43210, USA}


\author{Felix Kling}
\myaffiliation{Deutsches Elektronen-Synchrotron DESY, Notkestr. 85, 22607 Hamburg, Germany}

\author{Steven Linden}
\myaffiliation{Brookhaven National Laboratory, Upton, NY 11973, USA}

\author{Toni~M\"akel\"a}
\myaffiliation{Department of Physics and Astronomy, University of California, Irvine, CA 92697-4575, USA}

\author{Kostas Mavrokoridis}
\myaffiliation{University of Liverpool, Liverpool L69 3BX, United Kingdom}

\author{Josh McFayden}
\myaffiliation{Department of Physics \& Astronomy, University of Sussex, Sussex House, Falmer, Brighton, BN1 9RH, United Kingdom}

\author{Hidetoshi Otono}
\myaffiliation{Kyushu University, Nishi-ku, 819-0395 Fukuoka, Japan}

\author{Juan Rojo}
\myaffiliation{\mbox{Department of Physics and Astronomy, VU Amsterdam, 1081 HV Amsterdam, The Netherlands}}
\myaffiliation{\mbox{Nikhef Theory Group, Science Park 105, 1098 XG Amsterdam, The Netherlands}}

\author{Dennis Soldin}
\myaffiliation{\mbox{Department of Physics and Astronomy, University of Utah, Salt Lake City, UT 84112, USA}}

\author{Anna Stasto}
\myaffiliation{Department of Physics, Penn State University, University Park, PA 16802, USA}

\author{Sebastian Trojanowski}
\myaffiliation{National Centre for Nuclear Research, Pasteura 7, Warsaw, PL-02-093, Poland}

\author{Matteo Vicenzi}
\myaffiliation{Brookhaven National Laboratory, Upton, NY 11973, USA} 

\author{Wenjie Wu\vspace{1.0in}}
\myaffiliation{Department of Physics and Astronomy, University of California, Irvine, CA 92697-4575, USA}

\begin{abstract}
\vspace*{0.2in}
The Forward Physics Facility (FPF) is a proposal developed to exploit the unique scientific potential made possible by the intense hadron beams produced in the far-forward direction at the high luminosity LHC (HL-LHC).
Housed in a well-shielded cavern 627\,m from the LHC interactions, the facility will enable a broad and deep scientific programme which will greatly extend the physics capability of the HL-LHC. 
Instrumented with a suite of four complementary detectors -- FLArE, FASER$\nu$2, FASER2 and FORMOSA -- the FPF has unique potential to shed light on neutrino physics, QCD, astroparticle physics, and to search for dark matter and other new particles.
This contribution describes some of the key scientific drivers for the facility, the engineering and technical studies that have been made in preparation for it, the design of its four complementary experiments, and the status of the project's partnerships and planning.

\end{abstract}

\maketitle
\clearpage

\section*{INTRODUCTION AND SCIENTIFIC CONTEXT
\label{sec:intro}}
\vspace{-2mm}

\pagestyle{plain}
\setcounter{page}{1} 

High-energy colliders have enabled many groundbreaking discoveries since they were first constructed about 60 years ago.  As the latest example, the Large Hadron Collider (LHC) at CERN has been the center of attention in particle physics for decades. 
Despite this, the physics potential of the LHC is far from being fully explored, because the large detectors at the LHC are blind to collisions that produce particles in the forward direction, along the beamline. 
These forward collisions are a treasure trove of physics, providing the only way to study TeV neutrinos produced in the lab and unique opportunities to discover and study dark matter (DM) and other new particles beyond the Standard Model of particle physics.  

\begin{wrapfigure}{R}{0.43\textwidth}
\vspace*{-0.5in}
  \begin{center}
    \includegraphics[width=0.43\textwidth]{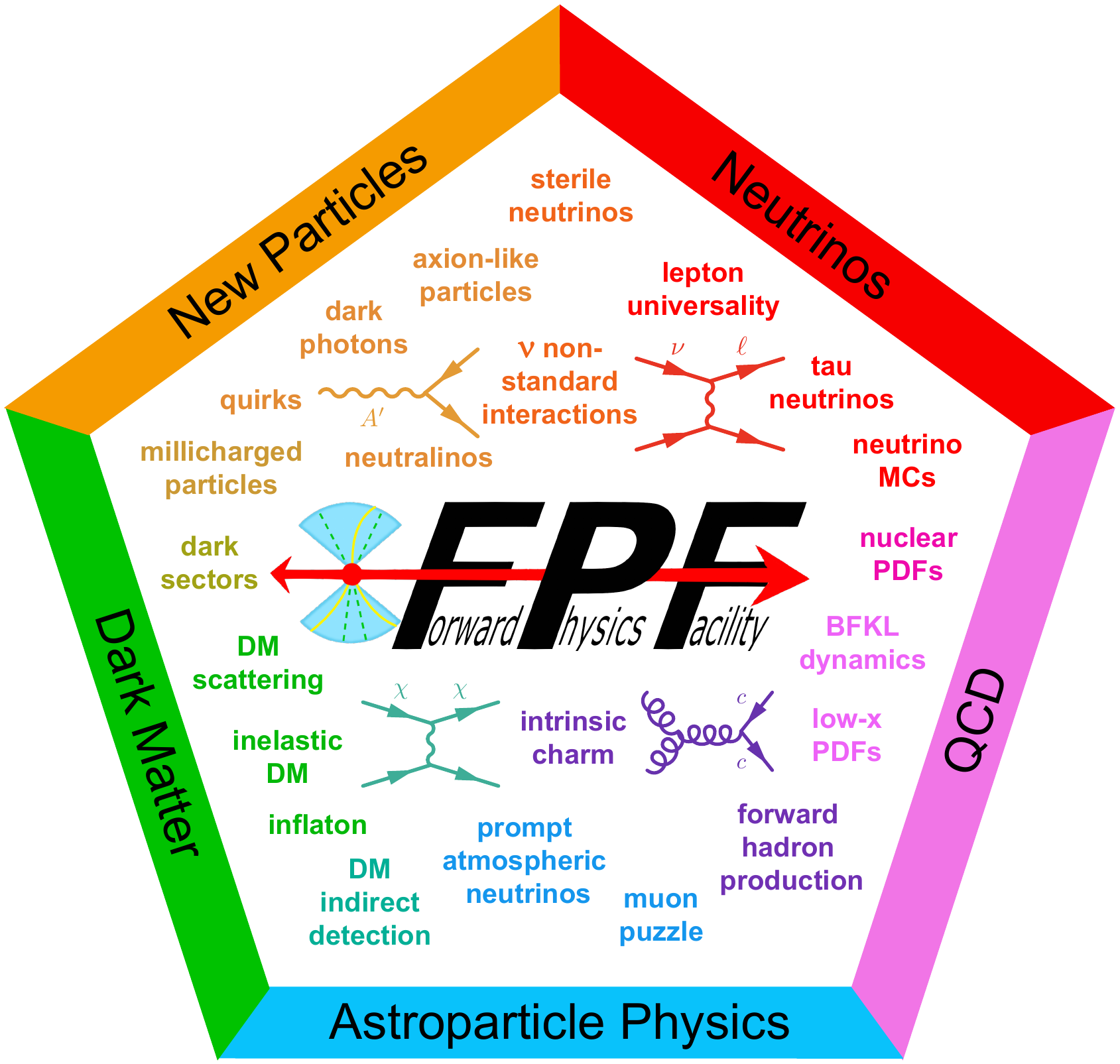}
\caption{The rich physics program at the FPF spans many topics and frontiers.
\label{fig:pentagon}}
  \end{center}
\vspace*{-0.3in}
\end{wrapfigure}

The Forward Physics Facility (FPF) is a proposal for a new underground cavern at CERN to house a suite of experiments during the High-Luminosity LHC (HL-LHC) era.
Its four complementary experiments -- FLArE, FASER$\nu$2, FASER2 and FORMOSA -- will cover the blind spots of the existing LHC detectors and are required if the LHC is to fully realize its physics potential. 
The physics program of the FPF is broad and deep; see \cref{fig:pentagon}.  The FPF can discover a wide variety of new particles that cannot be discovered at fixed target facilities or other LHC experiments. 
In the event of a discovery, the FPF, with other experiments, will play an essential role in determining the precise nature of the new physics and its possible connection to the dark universe.  In addition, the FPF is the only facility that will be able to detect millions of neutrinos with TeV energies, enabling precision probes of neutrino properties for all three flavors. 
These neutrinos will also sharpen our understanding of proton and nuclear structure, enhancing the power of new particle searches at ATLAS and CMS, and enabling IceCube, Auger, KM3NeT and other astroparticle experiments to make the most of the new era of multi-messenger astronomy. 

\vspace{-2mm}
\section*{PHYSICS OBJECTIVES
\label{sec:physics}}
\vspace{-2mm}

\noindent \textbf{Documentation:} The science case for the FPF has been developed in eight dedicated FPF meetings. The physics opportunities have been summarized in an 80-page review~\cite{Anchordoqui:2021ghd} and a more comprehensive 430-page White Paper~\cite{Feng:2022inv}, written and endorsed by 400 physicists. A recent summary written for the 2024-2026 European Particle Physics Strategy Update can be found in Ref.~\cite{Adhikary:2024nlv}. In the following, we present a few highlights of this broad program. \medskip

\noindent \textbf{Neutrino Physics at TeV Energies:} The LHC is the source of the most energetic neutrinos produced in a controlled laboratory environment, generating intense and strongly collimated beams of neutrinos and anti-neutrinos of all flavors in the forward direction. Although known since the 1980s~\cite{DeRujula:1984pg}, this has only recently been taken advantage of by the FASER~\cite{FASER:2022hcn} and SND@LHC~\cite{SNDLHC:2022ihg} experiments, which observed the interactions of collider neutrinos for the first time in 2023~\cite{FASER:2023zcr, SNDLHC:2023pun}. By the end of LHC Run~3 in 2026, the collaborations expect to detect approximately $10^4$ neutrinos. 

\begin{figure}[thb]
\centering
\includegraphics[width=0.325\textwidth]{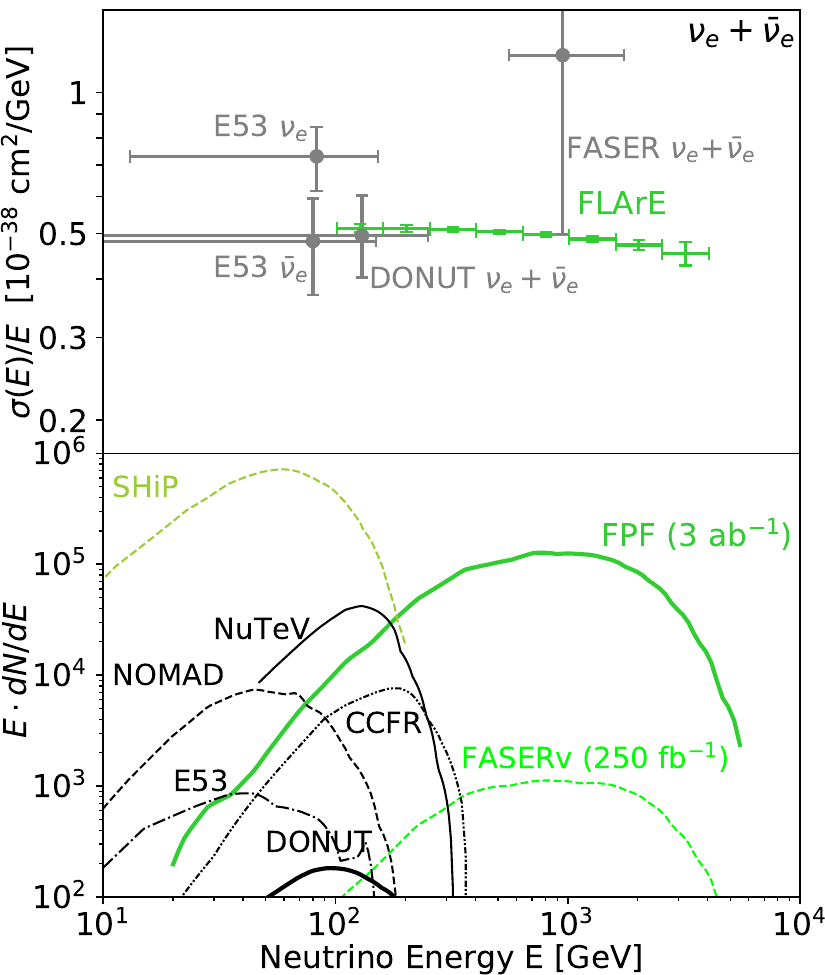}
\includegraphics[width=0.325\textwidth]{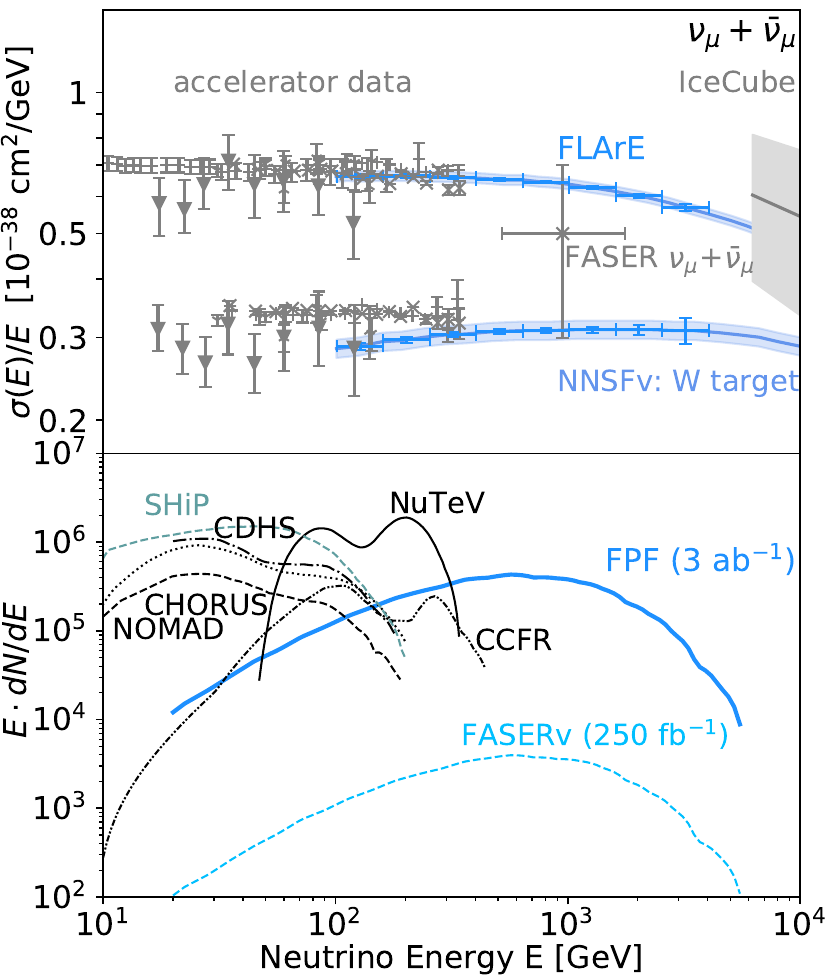}
\includegraphics[width=0.325\textwidth]{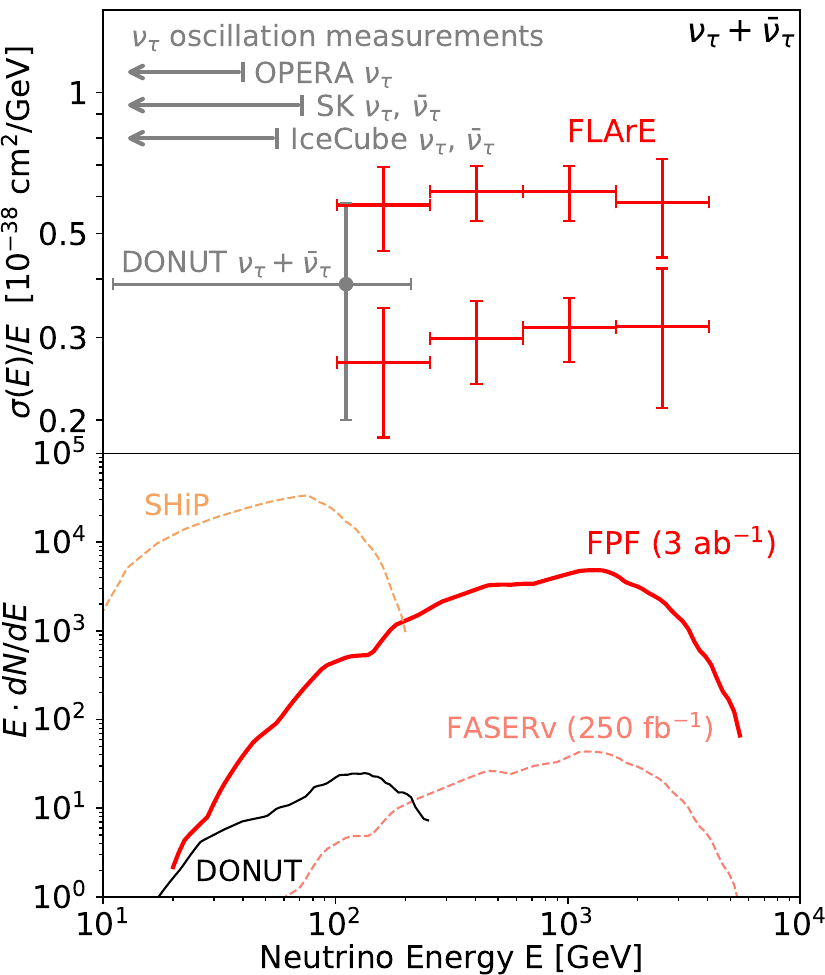}
\caption{\textbf{Neutrino yields and cross sections at the FPF.} The expected precision of FLArE measurements of neutrino interaction cross sections (top, statistical errors only) and the spectra of neutrinos interacting in the FPF experiments (bottom) as a function of energy, for electron (left), muon (middle), and tau (right) neutrinos. Separate measurements of muon and tau neutrinos and anti-neutrinos can be performed using muons passing through the FASER2 spectrometer. Existing data from accelerator experiments~\cite{ParticleDataGroup:2020ssz}, IceCube~\cite{IceCube:2017roe}, and the recent FASER$\nu$ result~\cite{FASER:2024hoe} are also shown, together with the prospects for SHiP~\cite{Ahdida:2023okr}.}
\label{fig:SM_neutrino}
\end{figure}

With larger detectors and higher luminosities, the FPF experiments are projected to detect $10^5$, $10^6$, and $10^4$ electron, muon, and tau neutrino interactions, respectively --- a hundred times the yields at existing experiments. The bottom panels of \cref{fig:SM_neutrino} show the expected energy spectra of interacting neutrinos at the FPF.  The collider neutrino energy spectrum peaks at $\sim \tev$ energies, never probed by previous measurements. This large event rate enables precision measurements for all flavors, and the first-ever distinction of tau neutrinos and anti-neutrinos. The top panel depicts the expected precision of the neutrino-nucleon charged-current scattering cross sections for all flavors. While the low energy ($\lesssim$\,100\,GeV)  region is well-constrained by accelerator neutrino experiments, and the very high energy muon neutrino cross section by IceCube using atmospheric neutrinos (with large uncertainties), FPF makes precise measurements at intermediate energies. 
\medskip

\begin{wrapfigure}{t}{0.53\textwidth}
  \vspace{-6mm}
  \centering \includegraphics[width=0.49\textwidth]{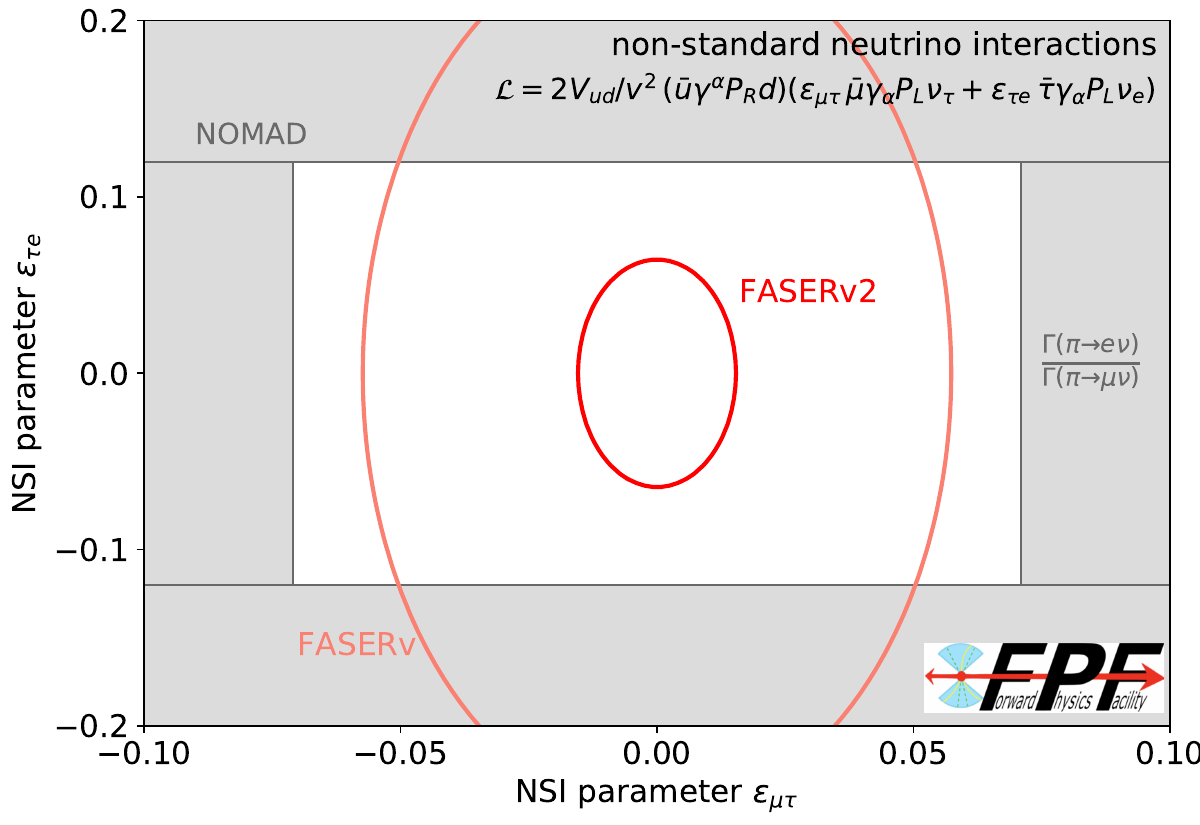}
  \vspace{-0.5cm}
  \caption{\textbf{Precision tau neutrino studies at the FPF.} The projected sensitivity of FASER$\nu$2 to neutrino NSI parameters violating lepton flavor universality~\cite{Kling:2023tgr}. }
  \label{fig:tau_neutrino}
  \vspace{-0.4cm}
\end{wrapfigure}

\noindent \textbf{Tau Neutrino Precision Measurements:} While previous experiments have identified only a few handfuls of tau neutrino interactions, FASER$\nu$2 will detect thousands. These measurements will provide the first definitive observation of anti-tau neutrinos and open up a
new window to an era of tau neutrino precision studies at TeV energies. In particular, these observations will enable tests of lepton flavor universality. Deviations from
universality may be parameterized, for example, by neutrino non-standard interactions
(NSI)~\cite{Falkowski:2021bkq}: an example of FASER$\nu$2’s unique sensitivity to probe NSI parameters associated with tau neutrinos is shown in \cref{fig:tau_neutrino}. \medskip 

\begin{figure}[t]
\centering
\includegraphics[width=0.8\textwidth]{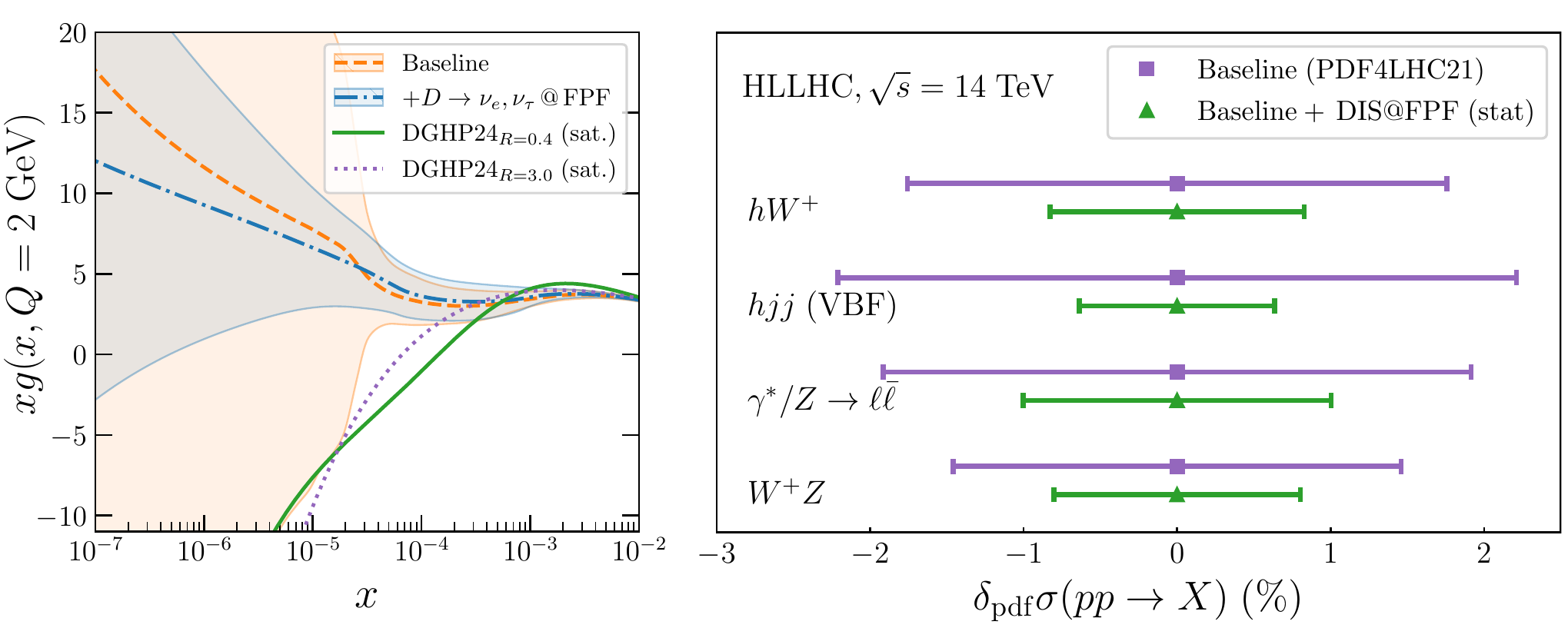}
\caption{Left: Impact of neutrino flux measurements at the FPF on the small-$x$ gluon PDF. Right: Reduction of the PDF uncertainties on Higgs- and weak gauge- boson cross sections at the HL-LHC, enabled by neutrino DIS measurements at the FPF.}
\label{fig:PDF}
\end{figure}

\vspace*{1.5cm}
\noindent \textbf{FPF as a Neutrino-Ion Collider:} The large number of neutrino events at the FPF will enable multi-differential measurements of their interaction cross-section, allowing to constrain proton and nuclear parton distribution functions (PDFs). The kinematics of these deep-inelastic scattering (DIS) events are parametrized in terms of momentum fractions $x \sim Q^2/(2 m_\text{nuc} E_\nu)$, with $Q > 2~\gev$ being the momentum transfer and $m_\text{nuc}$ the nucleon mass. Hence, the roughly ten times larger neutrino energies at the LHC, compared to previous accelerator experiments, probe roughly ten times smaller values of $x$. Dedicated projections demonstrated that $\mathcal{O}(10^6)$ muon-neutrino interactions at the FPF will provide stringent constraints on the proton PDFs~\cite{Cruz-Martinez:2023sdv}. These charge current DIS measurements at the FPF complement neutral current DIS measurements at the planned Electron-Ion Collider (EIC)~\cite{AbdulKhalek:2021gbh} as well as the proposed Large Hadron Electron Collider (LHeC)~\cite{LHeC:2020van}, while probing complementary flavor combinations. For example, the muon charge identification and $D$-meson tagging capabilities at the FPF allow selecting processes like $\nu s \to \mu^- c$ and $\bar\nu \bar s \to \mu^+ \bar c $, enabling distinction of specific initial state flavors, including the relatively poorly constrained strange quark content of the proton.

The impact of DIS neutrino measurements at the FPF on the traditional HL-LHC program is twofold. On one hand, FPF-constrained PDFs improve the precision of theoretical predictions for core processes at the HL-LHC, such as Higgs, Drell-Yan, and di-boson production (see right panel of Fig.~\ref{fig:PDF}), making the measurements of these cross sections at ATLAS and CMS more sensitive to physics beyond the Standard Model (BSM)~\cite{Cruz-Martinez:2023sdv}. On the other hand, BSM signals present in high-$p_T$ tails at the HL-LHC may inadvertently be reabsorbed in a PDF fit. The introduction of the FPF data in a PDF fit breaks the degeneracy between QCD and BSM effects, further enhancing the LHC discovery prospects~\cite{Hammou:2024xuj}. \medskip

\noindent \textbf{Unique Probe of small-$x$ QCD:} The LHC neutrino fluxes are sensitive to forward light and heavy hadron production in $pp$ collisions~\cite{Kling:2021gos, Buonocore:2023kna}. Both high-energy electron neutrinos and tau neutrinos originate primarily from charm hadrons, produced mainly via the fusion of a gluon carrying a large momentum fraction $x\sim 1$ with another carrying a very small momentum fraction $x \sim 4 m_c^2 / s \sim 10^{-7}$. For comparison, measurements of forward $D$-meson production at LHCb constrain the gluon PDF only down to $x\sim 10^{-5}$~\cite{Gauld:2016kpd, Zenaiev:2019ktw}. By exploiting observables where theory uncertainties cancel out, such as the ratio of electron and tau neutrino event rates, FPF measurements can pin the gluon PDF down as low as $x\sim 10^{-7}$~\cite{Rojo:2024tho}, as shown in the left panel of Fig.~\ref{fig:PDF}. Such measurements inform the study of novel QCD dynamics at small-$x$, a region where non-linear and BFKL-like effects are expected to dominate. This is highlighted by the DGHP24 predictions~\cite{Duwentaster:2023mbk} for the gluon PDF based on saturation (recombination) effects. Constraints on the small-$x$ gluon PDF will be instrumental to inform FCC-hh cross sections, since at $\sqrt{s}=100$ TeV even Higgs and gauge boson production become ``small-$x$'' processes, with potentially large corrections from BFKL resummation~\cite{Bonvini:2018vzv,Rojo:2016kwu}. These constraints on small-$x$ QCD are also relevant for astroparticle physics, as further discussed below. One should note that only LHC neutrino experiments can access this small-$x$ region crucial for probing QCD and astroparticle physics processes. E.g. beam dump experiments, such as SHiP~\cite{SHiP:2015vad}, involve neutrinos of much lower energy (tens of GeV), and access only $x> 10^{-2}$ both in production and scattering.
\medskip

\begin{figure}[htb]
\centering
\includegraphics[width=0.82\textwidth]{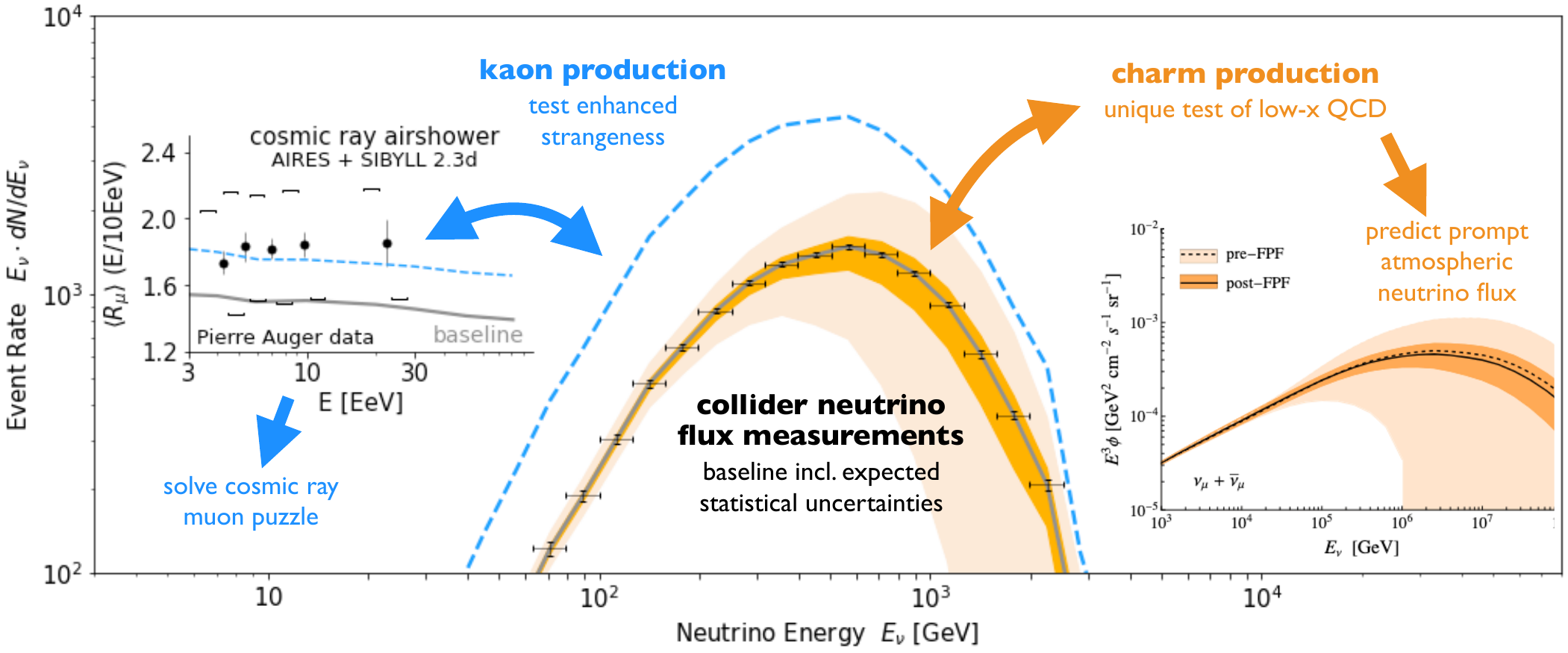}
\caption{\textbf{Astroparticle physics at collider neutrino experiments.} The central part of the figure shows the expected energy spectrum of interacting electron neutrinos in the FLArE detector at the FPF (solid gray curve) and expected statistical uncertainties (black error bars). The left part considers a model of strangeness enhancement introduced in Ref.~\cite{Anchordoqui:2022fpn}: a scenario that can resolve the discrepancy in the dimensionless muon shower content $R_\mu$ for air shower data would lead to sizable changes of the neutrino energy spectrum at the FPF and can therefore be tested. The right side of the figures shows how FPF data will reduce PDF uncertainties (orange band) on the prompt neutrino flux $\Phi$ as a function of $E_\nu$~\cite{Bai:2022xad}.}
\label{fig:Spectrum}
\end{figure}

\noindent \textbf{Impact for Astroparticle Physics:} LHC neutrino flux measurements will significantly boost astroparticle physics by validating and improving models of forward particle production.  This is essential for understanding particle production in extreme astrophysical environments and cosmic-ray interactions with the atmosphere.  Collider neutrino experiments offer a unique opportunity to test these models at comparable energies under controlled conditions. For instance, they will shed light on the cosmic ray muon puzzle --- a long-standing tension between measured and predicted muon counts in high-energy cosmic-ray air showers~\cite{PierreAuger:2016nfk, Soldin:2021wyv}. Thorough analyses show that an enhanced forward strangeness production could resolve this discrepancy~\cite{Albrecht:2021cxw}: this scenario can be tested using the LHC neutrino flux as show in Fig.~\ref{fig:Spectrum}. Additionally, the figure shows how collider data will reduce uncertainties in the prompt atmospheric neutrino flux~\cite{Gauld:2015kvh} --- arising from charmed hadron decays in cosmic-ray collisions --- which is the dominant background in astrophysical neutrino searches above a few 100~TeV~\cite{Bai:2022xad}. By measuring these limiting systematic effects the FPF will significantly enhance astrophysical neutrino studies and multi-messenger astronomy. \medskip

\noindent \textbf{Dark Matter and Dark Sector Mediators:} The nature of DM stands out as one of the foremost motivations for BSM physics. A generic, compelling possibility is that DM is part of a dark sector, feebly coupled to the SM by a mediator particle via portal interactions. If the mediator is the lightest state in the dark sector, it decays to SM particles through this interaction. 

\begin{wrapfigure}{l}{0.52\textwidth}
\centering
\vspace{-0.5cm}
\includegraphics[width=0.49\textwidth]{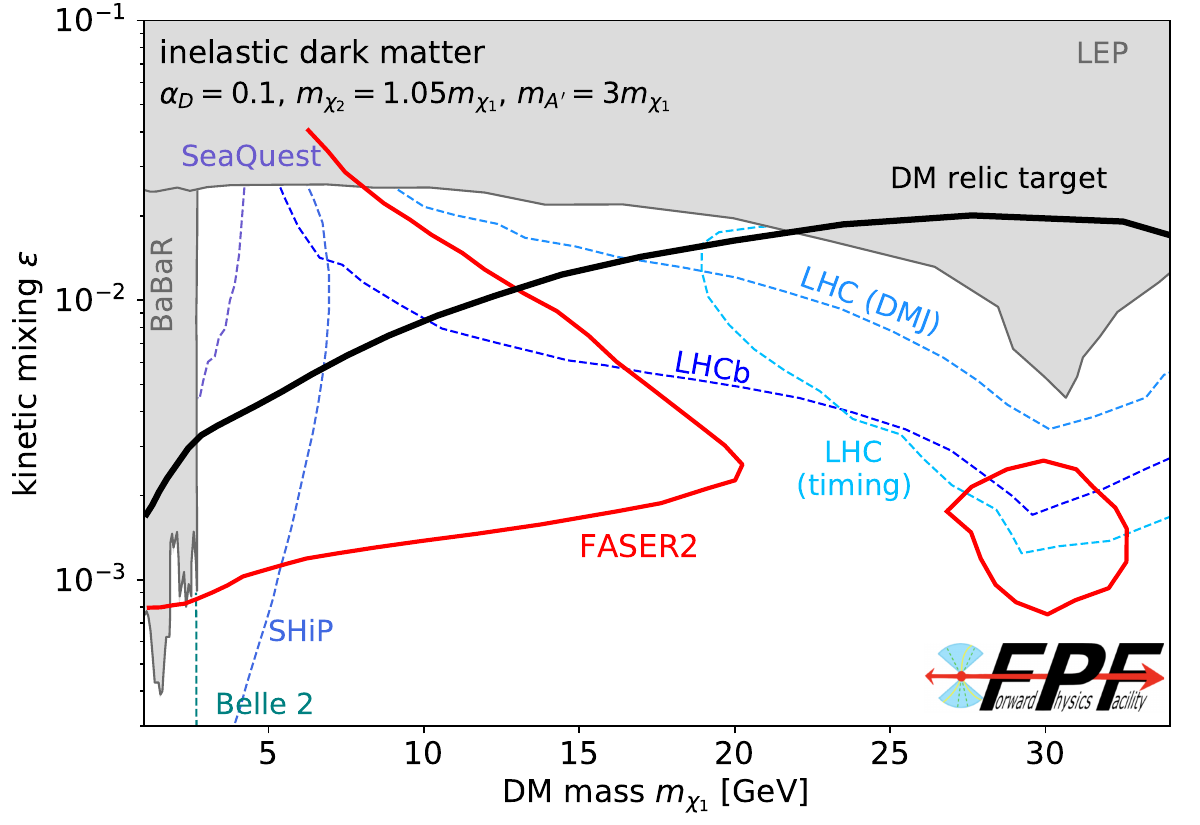}
\vspace{-0.3cm}
\caption{\textbf{Inelastic dark matter searches at the FPF}. The discovery potential of FASER2 and other experiments for heavy inelastic DM with masses up to tens of GeV~\cite{Berlin:2018jbm}. The reach of FASER2 extends beyond all other searches, including those at LHC and beam dump experiments, and covers the cosmologically-favored parameter space corresponding to the thermal relic target.}
\vspace{-0.3cm}
\label{fig:BSM_DM}
\end{wrapfigure}

The powerful capability of the FPF to search for a broad spectrum of long-lived particles was established in a multitude of publications and summarized in Ref.~\cite{Feng:2022inv}. Notably, this includes all benchmarks models discussed in the context of the Physics Beyond Colliders initiative~\cite{Alemany:2019vsk}: dark photons, dark Higgs, heavy neutral leptons, and axion-like particles.  In the event of a long-lived particle discovery, multiple experiments with complementary capabilities will be required to determine the fundamental properties of the new state (i.e., its mass, lifetime, spin, and couplings) and its connection to the dark universe. The FPF experiments will play an essential role in this endeavor.

Furthermore, there are well-motivated DM scenarios featuring a rich dark sector that can be uniquely probed by the FPF. For instance, \cref{fig:BSM_DM} illustrates the expected sensitivity of FASER2 to inelastic DM (iDM). In this model, an excited dark sector state decays into a lighter DM particle plus a visible final state~\cite{Berlin:2018jbm}. 
Such states, if heavier than a few GeV, are beyond kinematic threshold of beam dump experiments, but the high energy at the LHC provides large production rates, and the sensitivity of the FPF to highly-displaced decays enables a unique exploration of new regions in the parameter space, beyond the reach of the existing large LHC detectors. In this scenario, FASER2 will be able to decisively test a broad domain of parameter space where DM is produced in the early universe through thermal freeze-out.

Alternatively, DM may be significantly lighter than the mediator produced in $pp$-collisions at the LHC, which then decays ``invisibly'' to DM particle pairs. This yields a large flux of DM particles, detected via its scattering off electrons and nuclei in the FLArE and FASER$\nu$2 detectors, which can detect DM scattering in the relativistic regime, thus complementing traditional underground DM direct detection experiments~\cite{Batell:2021blf, Batell:2021aja}.
\medskip

\begin{wrapfigure}{t}{0.52\textwidth}
\centering
\vspace{-6mm}
\includegraphics[width=0.49\textwidth]{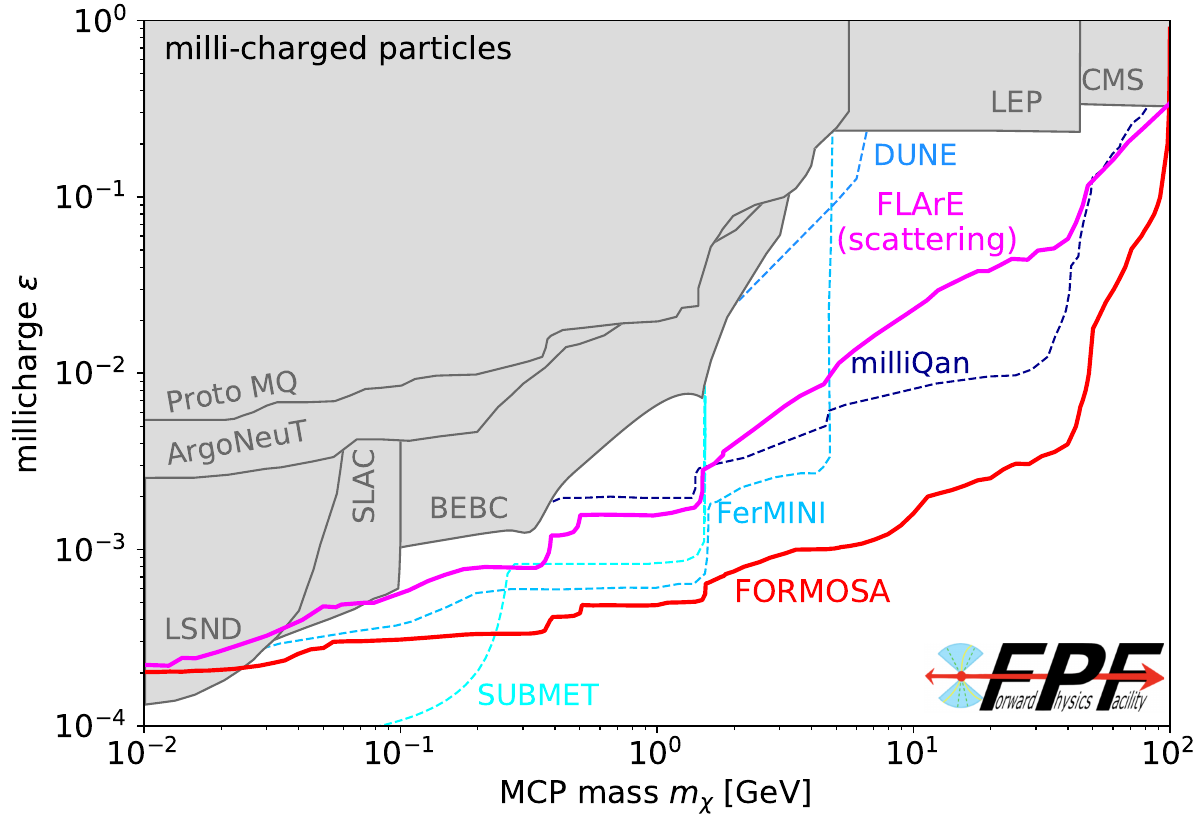}
\caption{\textbf{New particle searches at the FPF.} The discovery reach of FORMOSA and FLArE for millicharged particles~\cite{Foroughi-Abari:2020qar, Kling:2022ykt}. }
\vspace{-4mm}
\label{fig:BSM_newparticles}
\end{wrapfigure}

\noindent \textbf{Millicharged Particles:}  The prospects for millicharged particle (mCP) searches at the FPF are shown in \cref{fig:BSM_newparticles}. Such particles provide an interesting BSM physics target, both for their possible implications for charge quantization and as a candidate for a strongly interacting sub-component of DM. FORMOSA, a proposed scintillator-based experiment at the FPF, will have world-leading sensitivity to mCPs~\cite{Foroughi-Abari:2020qar}. Leveraging the high energy of the LHC collisions and the enhanced mCP production rate in the forward region, it will provide the most sensitive probe in the 100 MeV–100 GeV mass range compared to existing bounds and projections from several ongoing or proposed experiments. \medskip

\noindent \textbf{Other Opportunities for New Particle Searches:}  The many experimental signatures and broad range of BSM particle masses that can be probed at the FPF, from MeV to TeV scales, provide the foundation for a broad BSM physics program addressing fundamental questions in particle physics in a manner complementary to other existing and proposed facilities.  In addition to decays of long-lived particles, further signatures are provided by muon-philic particles produced in muon scattering in the neutrino detectors~\cite{Ariga:2023fjg, Batell:2024cdl, MammenAbraham:2025gai}, light axions converting to photons in the magnetic fields of FPF experiments~\cite{Kling:2022ehv}, and heavy quirks leading to delayed signals~\cite{Feng:2024zgp}. Additionally, the FPF will search for new physics in the neutrino sector, including searches for new neutrino-philic particles~\cite{Kling:2020iar, Kelly:2021mcd}; sterile neutrinos induced oscillations; anomalous electromagnetic properties of neutrinos~\cite{MammenAbraham:2023psg}; neutrino NSIs~\cite{Ismail:2020yqc}; and neutrino trident production~\cite{Altmannshofer:2024hqd}.

\section*{The Facility}

The FPF facility has been studied within the context of the CERN Physics Beyond Colliders effort over the last four years, with technical studies by CERN experts detailed in Refs.~\cite{PBCnote,PBCnote2,vibration-note}. 
The work has benefited from the deep experience at CERN in designing and implementing many similar large underground facilities, particularly the recent HL-LHC underground works close to ATLAS/CMS. Many of the same technical solutions can be adopted for the FPF, and lessons learned taken into account. \medskip

\begin{figure}[b]
\centering
\includegraphics[width=\textwidth,trim=0 200 0 30,clip]{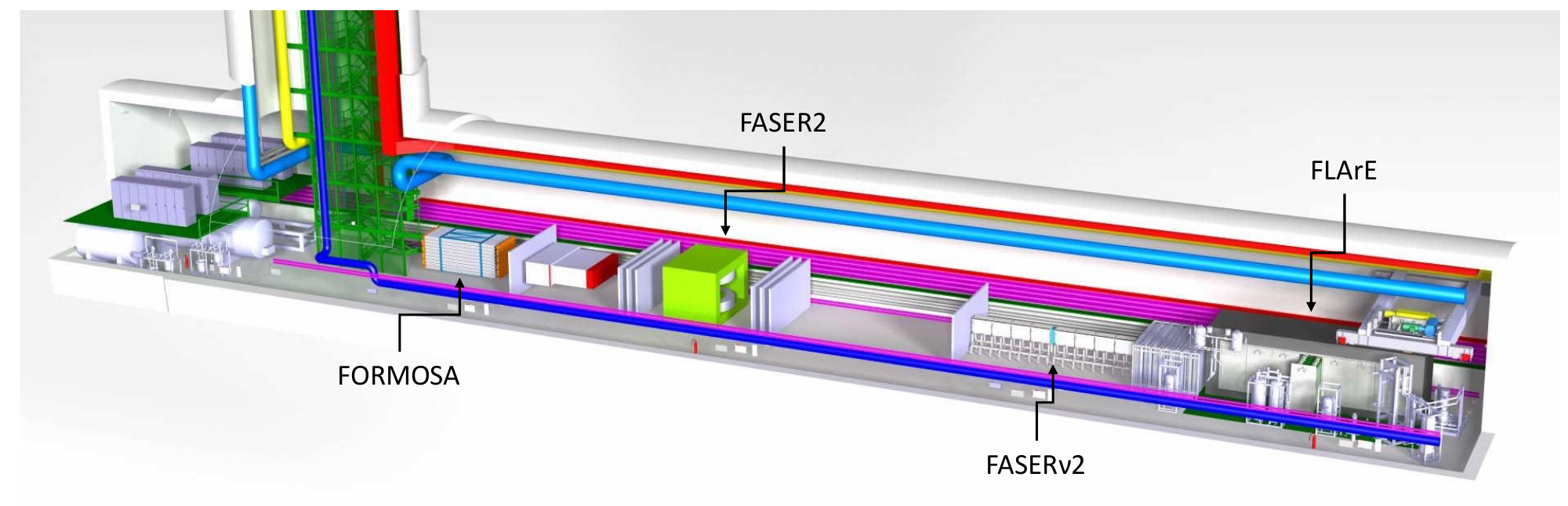}
\caption{The baseline layout of the FPF facility, showing the four proposed experiments and the large infrastructure. 
}
\label{fig:facility-layout}
\end{figure}

\noindent \textbf{Site Selection and Cavern Design:} A site optimization to find the best location for the FPF facility was carried out. This identified an optimal site 627~m west of the ATLAS IP (IP1), on CERN land in France. 
The baseline facility is shown in~\cref{fig:facility-layout} and includes a 75~m-long, 12~m-wide underground cavern, as well as an 88~m-deep shaft and the associated surface building for access and services. The closest point between the underground cavern and the LHC tunnel is 10~m, following civil engineering (CE) and radiation protection requirements. \medskip

\noindent \textbf{Site Investigation and Geological Conditions:} In Spring 2023, a site investigation study was carried out where a 20~cm-diameter, 100~m-deep core was drilled at the proposed location of the FPF shaft. Analysis of the extracted core confirmed that the geology is good for the planned excavation works, and no show stoppers were identified. \medskip 

\noindent \textbf{Excavation Work and Vibrations:} Carrying out the FPF excavation work during beam operation will allow much more flexibility in the implementation schedule. To address any potential concerns that the works might adversely impact beam operations, the CERN accelerator group has carried out detailed studies of the effect of the expected vibration level, documented in Ref.~\cite{vibration-note}. The conclusion is that no vibration problems are foreseen, and the excavation can be carried out during beam operations. \medskip 

\noindent \textbf{Radiation Levels and Safety:} Being able to access the cavern during beam operations will be extremely valuable during detector installation, commissioning, maintenance tasks, and for future detector upgrades. FLUKA simulations have been used to assess the radiation level in the FPF cavern during beam operation~\cite{PBCnote}. These show that the expected radiation level will be low enough for personnel to access the cavern during beam operation, with some restrictions. Simulations of the expected neutron and high-energy hadron flux also predict no problems from radiation effects: neither to electronics nor related to detector ageing. 
\medskip

\noindent \textbf{Technical Infrastructure and Integration:} Integration studies have shown that the proposed experiments, including their main associated infrastructure, can be installed and fit into the baseline cavern.  Standard infrastructure and services that have been considered include cranes and handling infrastructure, electrical power, ventilation systems, fire/smoke safety, access, and evacuation systems.  
\medskip

\noindent \textbf{Preliminary Facility Costing:} A Class 4 costing for the CE work has been carried out in September 2024, based on similar work undertaken at CERN in the last decade and taking into account the findings of the site investigation.  The costing methodology has been cross checked by an external CE consultant. The cost estimate is 35~MCHF for the underground works, shaft, and surface buildings.  A detailed breakdown of these costs is given in Ref.~\cite{Adhikary:2024nlv}. The expected time for the CE works is 3~years.  A very preliminary costing of the required services is at the level of less than 10~MCHF, giving a total costing of the facility, including both CE and outfitting, of around 45~MCHF.
\medskip

\noindent \textbf{Muon Fluxes:} The expected muon background rate in the FPF has been estimated using FLUKA~\cite{fluka} simulations. These simulations include a detailed description of the infrastructure between IP1 and the FPF. For the LHC Run 3 setup, the simulations have been validated with experimental measurements~\cite{FASER:2018bac,SNDLHC:2023mib}.  For the HL-LHC configuration, FLUKA predicts a muon flux of 0.6~cm$^{-2}$~s$^{-1}$ close to the line of sight for a luminosity of $5 \times 10^{34}$ cm$^{-2}$ s$^{-1}$. In general, the expected muon rate is acceptable for the proposed experiments, although reducing the rate would be beneficial and several options for this are under study. 

\vskip-3mm
\section*{The FPF Experiments}

Four complementary experiments~\cite{Adhikary:2024nlv} have been designed for the facility\footnote{The AdvancedSND experiment which was also included in e.g. Ref.~\cite{Feng:2022inv} no longer forms part of the FPF proposal.}, 
based on both experience with operating pathfinder experiments (FASER, FASER$\nu$ and milliQan)~\cite{FASER:2023tle,FASER:2024bbl, FASER:2023zcr,FASER:2024hoe,Ball:2020dnx} and analysis of a set of site-specific simulations and engineering studies for the FPF~\cite{Salin:2927003,Vicenzi:2927282,FLArETechNote}.  While each of the individual experiments have particular strengths, they have been designed and laid out in the FPF such that they provide complementary information, even at the level of  individual interactions -- for example FASER2 will make the muon spectrometer measurements required for the full physics exploitation of FASER$\nu$2 and FLArE. 
While the baseline conceptual design of the four experiments is designed to be deliverable with mature, proven technologies, alternative options employing novel technologies are also under consideration either as alternatives or upgrades.
\medskip 


\noindent \textbf{FLArE:} FLArE is a modularized, liquid argon, time-projection chamber (TPC) designed as a multi-purpose detector for a wide range of energies.
It is motivated by the requirements of neutrino detection~\cite{Anchordoqui:2021ghd} and light dark matter searches~\cite{Batell:2021blf} and builds on the considerable investment in liquid noble gas detectors over the last decade (ICARUS, MicroBooNE,  SBND, ProtoDUNE, and various components of DUNE). 
Liquid argon as an active medium allows one to precisely determine particle identification, track angle, and kinetic energy from tens of MeV to many hundreds of GeV, thus covering both dark matter scattering and high-energy LHC neutrinos. 

A significant effort has been carried out \cite{Vicenzi:2927282,FLArETechNote} to define the detector geometry, cryogenics, and integration within the FPF cavern, as well as the complementarity with the FASER2 spectrometer. The resulting design is based on a single-wall $8.8~\mathrm{m} \times 2.0~\mathrm{m} \times 2.4~\mathrm{m}$ (inner dimensions) foam-insulated cryostat. The TPC is segmented in 21 modules, arranged in a $3 \times 7$ configuration. 
The total liquid argon fiducial (active) mass is approximately $10~\mathrm{tons}$ ($30~\mathrm{tons}$). 
Each TPC module ($1.0~\mathrm{m} \times 0.6~\mathrm{m} \times 1.8~\mathrm{m}$) is divided into two volumes by a central cathode, with an anode at either end, resulting in 42 separate $30~\mathrm{cm}$ drift volumes. 
The anode charge readout will be  pixelated. Simulations suggest that a $5~\mathrm{mm}$ pixel size will satisfy the spatial resolution requirements for track reconstruction and particle identification.
At a typical drift field of $500~\mathrm{V/cm}$, one expects $\sim20,000$ electrons per pixel from minimum-ionizing muons and corresponds to a 30:1 signal-to-noise ratio assuming a total electronic noise of 500 electron equivalent noise charge (ENC). 
An alternative TPC design based on the ARIADNE program \cite{Lowe:2023pfk}, using a single vertical-drift, dual-phase module with optical readout, is also being considered; this design offers potential cost savings and synergy with existing R\&D efforts, but presents challenges with increased diffusion and space charge.

The downstream side of the FPF cavern is reserved for some of the cryogenic infrastructure, including storage tanks for liquid argon, and nitrogen and a Turbo-Brayton LN2 condenser, which is a commercial unit that reduces the need to provide LN2 for cooling continuously. These facilities are kept away from the detectors to reduce noise or vibration. 

To improve hadronic energy containment and muon tagging, a magnetized hadron calorimeter/muon spectrometer is envisioned downstream of the TPC. A possible design based on the Baby MIND neutrino detector concept employed in the WAGASCI experiment~\cite{Hallsjo:2018mmo} is 
studied. It consists of magnetized iron plates interleaved with scintillator modules that measure the particle position and the curvature of the track along the assembly. FLArE is expected to see about 25-50 high-energy neutrino events/ton/$\ifb$ of collisions, providing the opportunity to measure the neutrino fluxes and cross-section for all three neutrino flavors. 

The synergy with the FASER2 magnetic spectrometer is expected to provide up to $45-55\%$ acceptance for high-energy muons, depending on the final FASER2 magnet design. \medskip


\noindent \textbf{FASER$\nu$2:} FASER$\nu$2 is a 20-ton neutrino emulsion-based detector.
It will perform precision tau neutrino measurements and heavy flavor physics studies, testing lepton universality in neutrino scattering and new physics effects, as well as providing important input to QCD and astroparticle physics.
An emulsion gel with silver bromide crystals of 200~nm diameter will be used, which provides an intrinsic position resolution of 50~nm allowing identification of heavy flavor particles produced in neutrino interactions, including tau leptons and charm and beauty particles. 

The design comprises 3300 emulsion layers interleaved with 2-mm-thick tungsten plates. The total volume of the tungsten target is 40 cm $\times$ 40 cm $\times$ 6.6 m, with a mass of 20 tons. 
The FASER$\nu$2 design also includes interface detectors which will enable a FASER2-FASER$\nu$2 global analysis and make measurement of the muon charge possible, a prerequisite for $\nu_\tau$/$\bar\nu_\tau$ separation. 

Proof of concept is provided by the FASER$\nu$~\cite{FASER:2019dxq} pathfinder experiment which provided the first evidence for neutrino interaction candidates produced at the LHC in 2021~\cite{FASER:2021mtu}, and the first measurements of the $\nu_e$ and $\nu_\mu$ interaction cross sections at TeV energies in 2024~\cite{FASER:2024hoe}.  A global analysis that links information from FASER$\nu$2 with the FASER2 spectrometer enables charge measurements of muons from tau decays, and thereby the detection of $\bar\nu_\tau$ for the first time.

In addition to the test of a mechanical prototype, other test samples were produced and exposed to the beam. With these, one can check the long-term performance of emulsion films to take data through a year without replacing emulsion films. One can also test a new type of photo-development solution, which increases the gain of chemical amplification, which can help maximize the readout speed of the emulsion detector. The analysis of the samples is ongoing.

The emulsion film production and its readout will be conducted at facilities in Japan. The capacity of the film production facility~\cite{Rokujo:2024gqw} is 1200 m$^2$ per year. The scanning system can read out $\sim$0.5 m$^2$ per hour, or 1,000 m$^2$ per year.  Emulsion detector analysis will be limited by the accumulated track density and become difficult above $10^6$ tracks/cm$^2$ with the current tracking algorithms. New tracking algorithms will be developed to tolerate the high track density, such as with a machine learning method.  To keep the accumulated track density at an analyzable level, the emulsion films will be replaced once per year.  \medskip


\noindent \textbf{FASER2:} FASER2 is a large-volume detector designed for sensitivity to a wide variety of models of BSM physics and for precise electron and muon reconstruction for neutrino measurements. It builds on experience gained from the operation of the existing FASER pathfinder experiment~\cite{FASER:2022hcn}.

\begin{figure}[bh]
  \centering
  \includegraphics[width=0.9\textwidth]{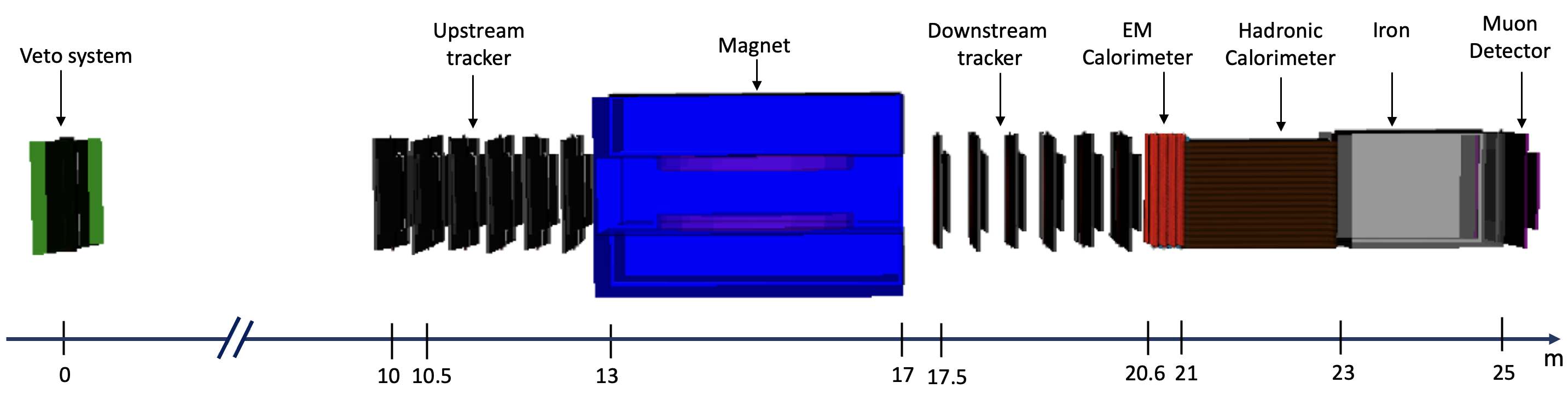}  
  \caption{Visualisation of the FASER2 detector, showing the veto system, uninstrumented 10\,m decay volume, tracker, magnet, electromagnetic calorimeter,  hadronic calorimeter, iron absorber and muon detector. The beam enters from the left.
  \label{fig:FASER2-Design}}
\end{figure}

The experiment design is dominated by a spectrometer with a large-volume superconducting dipole magnet. Investigations by KEK magnet experts, along with discussions with manufacturing experts at companies in Japan and in the UK, have demonstrated that this magnet design is feasible at an acceptable cost and lead time.  Alternative options are also being investigated to make use of industrial magnets with a smaller aperture (circular with 1.6 m diameter) and lower field strength ($\sim 1.25$--$2.50$ Tm)~\cite{Salin:2927003}. These magnets are commercially available, have performance within specification, and they are expected to be able to meet the physics objectives at reduced cost and lead time.

The baseline tracking detector design employs a SiPM and scintillating fiber tracker technology, based on LHCb's SciFi detector~\cite{Hopchev:2017tee}. This technology gives sufficient spacial resolution ($\sim 100\;\mu$m) at a significantly reduced cost compared to silicon detectors.  A lead-scintillator calorimeter design provides reconstruction of energy deposits from electrons and hadronic decay products of LLPs. Downstream is ${\cal O}(10)$ interaction lengths of iron with sufficient depth to absorb pions and other hadrons, followed by a scintillator detector for muon identification. The 10~m uninstrumented region upstream of the first tracking station (a $2.6 \times 1 \times 10$~m$^3$ cuboid) and downstream of the first veto station provides the decay volume for dark-sector particles.  This detector design, with  an intrinsic tracker resolution of 100~$\mu$m and a 2\,Tm integrated field strength, meets the performance requirements of a muon momentum resolution of approximately 2(4)\% for 1(5)~TeV muons~\cite{Salin:2927003}. 
Augmented tracking, with approximately $50 \mu m$ resolution, able to separate very closely spaced tracks near the beam axis is considered with HVCMOS sensors such as those used for LHCb’s Run 5 upgrade~\cite{Schmitz:2024kma,Scherl:2024fbe}.
\medskip


\noindent \textbf{FORMOSA:} FORMOSA meets the physics requirement for the FPF to have high sensitivity to millicharged particles~\cite{Foroughi-Abari:2020qar}. It will be technically similar to the milliQan pathfinder experiment~\cite{Haas:2014dda, Ball:2016zrp,milliQan:2021lne} at LHC Point 5, but with a larger active area and a better location with respect to the expected mCP flux, sensitivity for a given integrated luminosity is improved by a factor of ten. To meet the performance goals, plastic scintillator is chosen as the detection medium with the best combination of photon yield per unit length, response time, and cost~\cite{Haas:2014dda}. To maximize sensitivity to the smallest charges, each scintillator bar is coupled to a high-gain photomultiplier tube (PMT) capable of efficiently reconstructing the waveform produced by a single photoelectron (PE). 

The milliQan pathfinder experiment, operated during the Run~2 of the LHC, was successfully used to search for mCPs, proving the feasibility of such a detector~\cite{Ball:2020dnx}.
The feasibility of the experimental design has been demonstrated through the operation of the FORMOSA pathfinder/demonstrator operated behind FASER during Run~3 of the LHC.
Several studies of possible detector and physics backgrounds have also been performed~\cite{milliQan:2021lne,Foroughi-Abari:2020qar}, which together with the successful operation of the pathfinder experiments provide confidence that backgrounds are either negligible, or can be controlled.

\vskip-3mm
\section*{ Partnerships and Project Planning
\label{sec:internationalparticipation}}

The FPF and the forward physics detector collaborations will follow the best governance practices  established by other major collaborations such as ATLAS, CMS, DUNE, etc.  The scale of the FPF enterprise is much smaller than these  collaborations, and so suitable adjustments will be made to the governance and coordination practices. \medskip 

\noindent \textbf{Community:} Three of the proposed experiments, FASER2, FASER$\nu$2, and FORMOSA, have pathfinder projects that are already installed and running at the LHC: FASER, FASER$\nu$, and milliQan, respectively~\cite{FASER:2023tle,FASER:2024bbl, FASER:2023zcr,FASER:2024hoe,Ball:2020dnx}. 
Currently, the broader FPF experimental and theoretical community is approximately $\sim$400 strong, mostly from Europe, US, and Japan; see the list of contributors to and endorsers of Ref.~\cite{Feng:2022inv}.  
The collaborations behind the existing experiments as well as the broader community are accumulating considerable experience for this science and  will serve all experiments in the FPF.  \medskip

\noindent \textbf{Readiness:} The FPF project design is in pre-conceptual stage with a number of internal technical documents regarding cost, schedule, and facility integration ready for review.  A full conceptual design report along with  a management   structure under the guidance of CERN can be prepared in a short period of time. This structure will need to have a strong technical coordination team to construct the facility according to the scientific and engineering requirements, and also to install the detectors.  Further details on management and status of the collaboration can be found in Ref.~\cite{Adhikary:2024nlv}.  \medskip

\noindent \textbf{Budget:} A preliminary budget and schedule have been assembled for the FPF facility and the component experiments after discussions in successive FPF workshops. 
Detailed costs and a milestone schedule can be found in Ref.~\cite{Adhikary:2024nlv}; here we will provide a brief summary.  

The costs are divided into facility costs -- for civil construction and additional infrastructure -- plus the costs for the individual experiments.  The overall concept has undergone considerable improvement during 2024 due to engineering integration and improved experimental planning.  Underground space, cryogenic infrastructure, ventillation, safety, and experimental installation are all ready for conceptual design.  
The cost for the civil construction and the outfitting was provided by the CERN civil engineering group and the technical infrastructure groups, respectively.  They reviewed the  initial experimental  requirements for the needed location, underground space, and services, and performed a Class 4  estimate~\cite{CEcosting} of approximately ~45MCHF for the facility.  Additional cryogenic infrastructure mainly for the FLArE (liquid argon TPC) neutrino detector will require about ~4MCHF.   

The costs for the experimental program were assembled by the proponents.   The costs for FASER2 are dominated by the proposed magnetic spectrometer systems. For FASER$\nu$2, the costs are dominated by the production and handling of emulsion. FORMOSA costs of plastic scintillator and PMT based readout are well known.  FLArE costs are dominated by cryogenic systems and the cryostat. All of these are well known due to considerable experience at CERN, especially at
the Neutrino Platform.
The international division of scope for components for these projects is currently not well defined, and therefore we have assembled core costs without labour, overhead, contingency, and additional factors that need to be considered according to the rules of each national sponsor.   These costs will be refined and further improved as we proceed to the conceptual design. 

The total core costs for the experimental program has been assembled to be about 41MCHF.  The cost is not dominated by any single experiment or infrastructure costs.  A few additional comments are necessary to understand options for the full FPF facility and the attendant experiments.  
The cost for FASER$\nu$2 includes the cost of replacing the emulsion films 10 times. These costs could change over time or be absorbed in the costs of detector operations. The baseline  for FLArE is now a single walled foam insulated  cryostat with modular TPCs with pixel readout; this could change to other options (e.g. optical readout) after R\&D.  The cryostat/cryogenics and additional infrastructure design and costs clearly need to be coordinated and shared with CERN. This process of coordination has started only recently.

The  cost for  experiments  does not include engineering, labor, project management, contingency, and   the research support that will be needed.  Obviously, for an effort of this size, considerable support will also be needed by a collaboration for students, postdocs, travel, and R\&D.  We estimate the total size of the collaboration to range from 250 to 350 people with corresponding annual support from the national agencies. 
FPF and its experiments is a medium sized  acquisition for CERN and potential partner national sponsors. We expect to benefit from the vast experience at CERN regarding  review processes and resource organization.  

We have prepared an approximate funding profile using the current understanding of the cost estimates for components as a planning exercise. The funding profile calls for ramped up funding for the facility and experiments starting in 2026 and completing with detector commissioning in 2032 in time for initial data from Run 4; data from Run 4 will allow for scientific productivity and strengthening of the FPF community.   We have considered that any FPF construction must be coordinated with the HL-LHC, so that the civil construction and demands on personnel do not interfere with LHC operations.  The CERN radiation protection group has concluded that the FPF can be accessed during LHC operations with appropriate controls for radiation safety.  This will allow detector installation to proceed during Run 4. Lastly, planning and integration of the FPF program will require excellent technical coordination with leadership from the host laboratories. \medskip

\noindent \textbf{Sustainability:} The FPF detectors are concurrent experiments at the LHC. Since the particles they aim to detect are already being produced in LHC collisions, they do not require additional large amounts of energy to generate a beam, but instead maximize the physics that can be done with existing beams and energy consumption.\medskip

\noindent \textbf{Technology:} The FPF experiments have been designed such that the baseline performance is deliverable with proven technologies. The continuous possible access to the FPF cavern, even during LHC operations, means that upgrades to detectors are possible and that the FPF could perform the role of a test facility for novel detectors, including for future colliders. \medskip

\noindent \textbf{People and Skills:} The FPF will attract a large and diverse global community. As a mid-scale project composed of smaller experiments that can be realized on short
and flexible timescales, the FPF will provide a multitude of scientific and leadership opportunities
for early-career researchers. They will make important contributions from design and construction to commissioning and data analysis, often in a single graduate student lifetime. That cohort will be part of the skills pipeline for longer-timescale major projects. \medskip

\noindent \textbf{Timeliness:} To fully exploit the forward physics opportunities the FPF and its experiments should be ready for physics in the HL-LHC era as early as is achievable within Run 4. 


\acknowledgements

We gratefully acknowledge the invaluable support from the CERN Physics Beyond Colliders study group, who have contributed many technical studies related to the feasibility of the implementation of the FPF.
ST is supported by the National Science Centre, Poland, research grant No.~2021/42/E/ST2/00031. 
The work of LAA is supported by the U.S. National Science Foundation grant PHY-2412679.
The work of AJB is funded in part through STFC grants ST/R002444/1 and ST/S000933/1.
The work of BB is supported by U.S.~Department of Energy grant DE–SC-0007914.
The work of J.Bian, MVD, SL, MV, and WW is supported in part by Heising-Simons Foundation Grant 2022-3319.
The work of MC is supported in part by U.S.~Department of Energy grant DE-SC0009999. 
The work of JLF and TM is supported in part by U.S.~National Science Foundation grants PHY-2111427 and PHY-2210283. 
The work of JLF is supported in part by Simons Investigator Award \#376204, Heising-Simons Foundation Grants 2019-1179 and 2020-1840, and Simons Foundation Grant 623683.  
The work of CSH is supported in part by U.S.~Department of Energy Grant DE-SC0011726.
The work of FK is supported by the Deutsche Forschungsgemeinschaft under Germany's Excellence Strategy -- EXC 2121 Quantum Universe -- 390833306.
The work of JM is supported by the Royal Society grant URF\textbackslash R1\textbackslash201519 and STFC grant ST/W000512/1.
The work of JR is partially supported by NWO, the Dutch Research Council, and by the Netherlands eScience Center (NLeSC).
The work at BNL is under U.S.~Department of Energy contract No.~DE-SC-0012704.  


\bibliography{references}

\end{document}